\documentclass[10pt]{article}
\usepackage{jheppub}
\usepackage{axodraw2}

\newcommand{\ba}{\begin{eqnarray}}
\newcommand{\ea}{\end{eqnarray}}
\newcommand{\af}{a_f}
\newcommand{\cf}{c_f}
\newcommand{\nf}{n_f}
\newcommand{\str}{{\rm sTr}}
\newcommand{\la}[1]{\label{#1}}
\newcommand{\fig}{figure~}
\newcommand{\eq}{eq.~}
\newcommand{\se}{section~}

\newcommand{\nr}[1]{(\ref{#1})}

\newcommand{\ep}{\varepsilon}
\newcommand{\CA}{C_{\mathrm{A}}}
\newcommand{\CF}{C_{\mathrm{F}}}
\newcommand{\NA}{N_{\mathrm{A}}}
\newcommand{\TF}{T_{\mathrm{F}}}
\newcommand{\Nf}{N_{\mathrm{f}}}

\newcommand{\msbar}{{\overline{\mbox{\rm{MS}}}}}
\newcommand{\code}[1]{{\tt #1}}

\newcommand{\diagA}{
\begin{axopicture}(200,120)
\Gluon(0,60)(50,60){3}{5}
\Gluon(150,60)(200,60){3}{5}
\GluonCirc(100,60)(45,0){3}{40}
\SetColor{Red}
\BCirc(145,60){19}
\BCirc(55,60){19}
\BCirc(100,105){15}
\BCirc(100,15){15}
\end{axopicture}
}
\newcommand{\diagB}{
\begin{axopicture}(200,120)
\Gluon(0,60)(40,60){3}{4}
\Gluon(160,60)(200,60){3}{4}
\GluonArc(100,132)(60,217,323){3}{20}
\SetColor{Red}
\ECirc(100,60){60}
\BCirc(100,70){10}
\BCirc(70,80){10}
\BCirc(130,80){10}
\end{axopicture}
}
\newcommand{\diagC}{
\begin{axopicture}(200,120)
\Gluon(0,60)(50,60){3}{5}
\Gluon(148,60)(200,60){3}{5}
\GluonCirc(100,60)(45,0){3}{40}
\SetColor{Red}
\BCirc(55,60){19}
\BCirc(90,105){10}
\BCirc(130,95){10}
\BCirc(100,15){15}
\end{axopicture}
}
\newcommand{\diagD}{
\begin{axopicture}(200,120)
\Gluon(0,60)(52,60){3}{6}
\Gluon(150,60)(200,60){3}{5}
\GluonCirc(100,60)(45,0){3}{40}
\SetColor{Red}
\BCirc(145,60){19}
\BCirc(100,15){15}
\BCirc(100,105){15}
\SetColor{Black}
\Gluon(160.5,49.5)(132,68.5){3}{4}
\Gluon(127,72)(100,90){3}{4}
\end{axopicture}
}
\newcommand{\diagE}{
\begin{axopicture}(200,120)
\Gluon(0,60)(50,60){3}{5}
\Gluon(150,60)(200,60){3}{5}
\GluonCirc(100,60)(45,0){3}{40}
\SetColor{Red}
\BCirc(145,60){19}
\BCirc(55,60){19}
\BCirc(100,105){15}
\SetColor{Black}
\Gluon(160.5,49.5)(132,68.5){3}{4}
\Gluon(127,72)(100,90){3}{4}
\end{axopicture}
}
\newcommand{\diagF}{
\begin{axopicture}(200,120)
\Gluon(0,60)(40,60){3}{4}
\Gluon(160,60)(200,60){3}{4}
\GluonArc(100,132)(60,217,323){3}{16}
\Gluon(100,120)(100,82){3}{4}
\SetColor{Red}
\ECirc(100,60){60}
\BCirc(100,70){12}
\BCirc(129,77.5){9}
\end{axopicture}
}
\newcommand{\diagG}{
\begin{axopicture}(200,120)
\Gluon(0,60)(50,60){3}{5}
\Gluon(150,60)(200,60){3}{5}
\GluonCirc(100,60)(45,0){3}{40}
\Gluon(100,90)(100,30){3}{10}
\SetColor{Red}
\BCirc(145,60){19}
\BCirc(100,105){15}
\BCirc(100,15){15}
\end{axopicture}
}
\newcommand{\diagH}{
\begin{axopicture}(200,120)
\Gluon(0,60)(40,60){3}{4} \Gluon(160,60)(200,60){3}{4}
\SetColor{Red} \ECirc(100,60){60}
\SetColor{Black} \Gluon(50,93)(100,60){3}{7} \Gluon(100,60)(150,93){3}{7}
\SetColor{Red} \BCirc(100,60){40}
\SetColor{Black} \Gluon(74,30)(100,60){3}{4} \Gluon(100,60)(126,30){3}{4}
\SetColor{Red} \BCirc(100,60){20}
\end{axopicture}
}
\newcommand{\diagI}{
\begin{axopicture}(200,120)
\Gluon(0,60)(40,60){3}{4}
\Gluon(160,60)(200,60){3}{4}
\GluonArc(100,132)(60,217,248){3}{3}
\GluonArc(100,132)(60,292,323){3}{4}
\GluonCirc(100,75)(20,14){3}{17}
\SetColor{Red}
\ECirc(100,60){60}
\BCirc(77,77){12}
\BCirc(103,56){10}
\end{axopicture}
}
\newcommand{\diagJ}{
\begin{axopicture}(200,120)
\Gluon(0,60)(50,60){3}{5}
\Gluon(148,60)(200,60){3}{5}
\GluonCirc(100,60)(45,0){3}{40}
\Gluon(122,80)(77,96){3}{7}
\SetColor{Red}
\BCirc(55,60){19}
\BCirc(100,88){8}
\BCirc(135,85){14}
\end{axopicture}
}
\newcommand{\diagK}{
\begin{axopicture}(200,120)
\Gluon(0,60)(20,60){3}{2}
\Gluon(180,60)(200,60){3}{2}
\GluonArc(100,-10)(100,65,115){3}{11}
\GluonArc(80,100)(65,250,320){3}{9}
\GluonArc(120,100)(65,220,244){3}{3}
\GluonArc(120,100)(65,259,290){3}{4}
\SetColor{Red}
\BCirc(45,60){25}
\BCirc(155,60){25}
\BCirc(100,90){10}
\end{axopicture}
}

\title{Towards the five-loop Beta function for a general gauge group}

\preprint{TTP16-025\\\mbox{}\hfill IPPP/16/57\\\mbox{}\hfill DESY 16-117}

\author[a]{Thomas Luthe,}
\author[b]{Andreas Maier,}
\author[c]{Peter Marquard}
\author[d]{and York Schr\"oder}

\affiliation[a]{Institut f\"ur Theoretische Teilchenphysik, Karlsruhe Institute of Technology, Karlsruhe, Germany}
\affiliation[b]{Institute for Particle Physics Phenomenology, Durham University, Durham, United Kingdom}
\affiliation[c]{Deutsches Elektronen Synchrotron (DESY), Platanenallee 6, Zeuthen, Germany}
\affiliation[d]{Grupo de Fisica de Altas Energias, Universidad del Bio-Bio, Casilla 447, Chillan, Chile}

\emailAdd{thomas.luthe@kit.edu}
\emailAdd{andreas.maier@durham.ac.uk}
\emailAdd{peter.marquard@desy.de}
\emailAdd{yschroeder@ubiobio.cl}

\arxivnumber{1606.08662}

\abstract{We present analytical results for the $\Nf^4$ and $\Nf^3$ terms of the five-loop Beta function, for a general gauge group. While the former term agrees with results available from large-\/$\Nf$ studies, the latter is new and extends the value known for SU(3) from an independent calculation.}

\begin{document}
\maketitle

%
\section{Introduction}
\la{se:intro}

The Beta function of Quantum Chromodynamics (QCD) governs the behavior of the strong interactions, as the energy scale is varied. 
As such, it plays a central role in the Standard Model of elementary particle physics, and indeed in quantum field theory, establishing an example of an asymptotically free theory. 
Of utmost present importance, on the phenomenological side, high-precision QCD results are needed in order to take full advantage of the experimental program being undertaken at the LHC. 

Given this importance, much effort has been spent on evaluating the fundamental building blocks of our theories with the best possible precision. The QCD Beta function has been calculated at the one- \cite{Gross:1973id,Politzer:1973fx}, two- \cite{Caswell:1974gg,Jones:1974mm}, three- \cite{Tarasov:1980au,Larin:1993tp} and four-loop \cite{vanRitbergen:1997va,Czakon:2004bu} levels in the past; first 5-loop results have started to appear, such as for the QED case \cite{Baikov:2008cp,Baikov:2010je,Baikov:2012zm}; at 6 loops, presently only scalar theory is accessible \cite{Batkovich:2016jus}.

The QCD Beta function can be evaluated from a variety of field and vertex renormalization constants.
A simple choice that we shall adopt here are the ghost propagator and -vertex, as well as the gluon propagator.
This amounts to computing the three renormalization constants $Z_{cc}$, $Z_{ccg}$ and $Z_{gg}$, the latter one being by far the most complicated to evaluate, due to the number of terms in the bare gluon vertices (and, if considering a general covariant gauge with gauge parameter $\xi$, in the bare gluon propagator) as compared to bare ghost and fermion couplings and -propagators.

The paper is structured as follows. 
In \se\ref{se:setup}, we start by explaining our computational setup and by discussing classes of Feynman diagrams that contribute to different color structures.
Using some notation defined in \se\ref{se:notation}, we then present and discuss results in \se\ref{se:results}, before concluding in  \se\ref{se:conclu}.

%
\section{Setup}
\la{se:setup}

Let us now give some details on our setup, and make some technical remarks.
As mandatory for a high-order perturbative calculation, we employ a highly automatized setup based on the diagram generator \code{qgraf} \cite{Nogueira:1991ex,Nogueira:2006pq} and several own \code{FORM} \cite{Vermaseren:2000nd,Tentyukov:2007mu,Kuipers:2012rf} programs.
After applying projectors and performing the color algebra using the \code{color} package \cite{vanRitbergen:1998pn}, we introduce a common mass into all propagators, and expand deep enough in the external momenta \cite{Misiak:1994zw,vanRitbergen:1997va,Chetyrkin:1997fm} -- this step is justified, since we are interested in ultraviolet (UV) divergences only, which allows us to regularize the small-momentum behavior of each Feynman diagram at will, at the minor cost of one new (gluon mass) counterterm.
Keeping all potentially UV divergent structures and nullifying the external momenta, the resulting expansion coefficients can then be mapped onto a set of vacuum integral families, which at 5-loop level are labelled by 15 indices \cite{Luthe:2015ngq}. 
Let us note that, while the maximum number of positive indices (lines) is 12 in the highest integral sector, for the present calculation we need a maximum of 11 only.
The resulting sum of fully massive scalar 5-loop vacuum integrals is then reduced to master integrals by systematic use of integration-by-parts (IBP) identities \cite{Chetyrkin:1981qh} using a Laporta-type algorithm \cite{Laporta:2001dd} as implemented in \code{crusher} \cite{crusher}. 

The required set of 110 master integrals has been evaluated at 5 loops previously, using a highly optimized and parallelized setup \cite{Luthe:2015ngq,Luthe:2016sya} based on IBP reduction and difference equations \cite{Laporta:2001dd}, implemented in \code{C++} and using \code{Fermat} \cite{fermat} for fast polynomial algebra. 
While all difference equations and recurrence relations have been obtained exactly in the space-time dimension $d$, a high-precision numerical solution of the required integrals around $d=4-2\ep$ dimensions allows to access (by far) sufficiently high orders in $\ep$ for our masters. 
These results satisfy a number of nontrivial internal checks, and correctly reproduce all lower-order results as well as (the few) previously known 5-loop coefficients.
In addition, we have used the integer-relation finding algorithm \code{PSLQ} \cite{MR1489971} to find the analytic content of some of these numerical results, and to discover linear relations among others. 
This final step allows us to express our result in terms of Zeta values only.

\begin{figure}
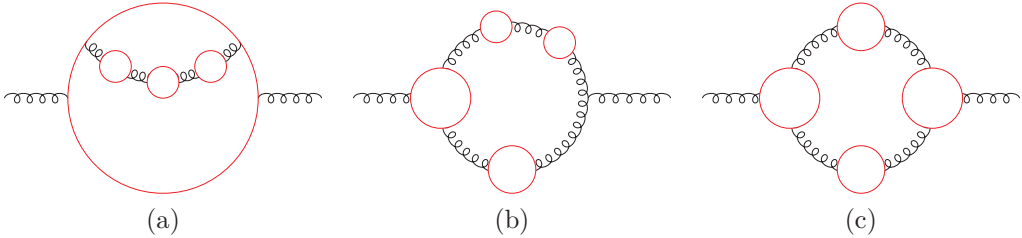

\SetScale{0.6}
\centerline{\begin{tabular}{ccc}
\diagB & \diagC & \diagA \\
(a) & (b) & (c)
\end{tabular}}
\caption{\label{fig:nf4}Sample 5-loop diagrams that contribute to different color structures at $\Nf^4$. Straight (red) and curly (black) lines denote quarks and gluons, respectively. We did not draw arrows on the fermion lines, since all combinations occur.}
\end{figure}

Turning now to the diagrammatic content of our calculation, at order $\Nf^4$, there are three distinct color structures that potentially contribute to the 5-loop Beta function. 
For the precise definition of these factors (such as $\cf$, $d_0$ etc.), we refer to the next section. 
Taking the gluon propagator as an example, \fig\ref{fig:nf4} lists one representative Feynman diagram for each of these color structures. 
Diagrams (a) and (b) contribute to $\cf$ and $1$ of \eq\nr{eq:result}, respectively. 
Diagrams of type (c) are proportional to $d_0$ individually; however, they cancel in the sum of diagrams, due to the structure of the two fermion loops making up $d_{33FF}$ (note that for the same reason, $d_0$ did not occur in the 3- and 4-loop Beta function coefficients either). 

\begin{figure}
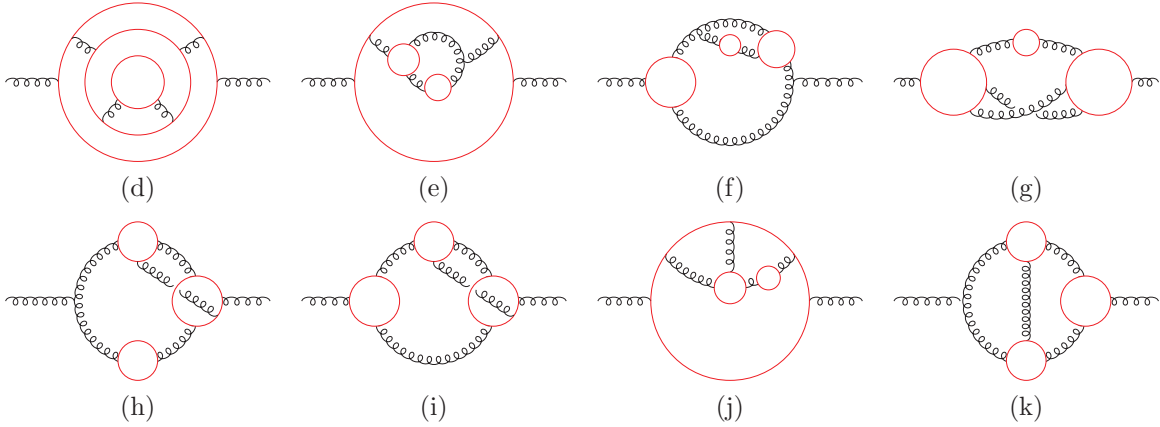

\SetScale{0.5}
\centerline{\begin{tabular}{cccc}
\diagH & \diagI & \diagJ & \diagK \\
(d) & (e) & (f) & (g) \\[2mm]
\diagD & \diagE & \diagF & \diagG \\
(h) & (i) & (j) & (k) 
\end{tabular}}
\caption{\label{fig:nf3}Sample 5-loop diagrams that contribute to different color structures at $\Nf^3$. The notation is as in \fig\ref{fig:nf4}. Only the diagram classes whose representatives are depicted in the first line need to be considered in practice. For details, see the main text.}
\end{figure}

In the same spirit, \fig\ref{fig:nf3} depicts representatives that contribute to distinct color structures at order $\Nf^3$.
Diagrams (d)--(g) contribute to $\cf^2$, $\cf$, $1$ and $d_1$ of \eq\nr{eq:result}, respectively. 
Diagrams of class (h) actually vanish individually after performing the color sums, while diagrams of class (i) are proportional to $d_4$ individually, but cancel in the sum. 
The last two classes of \fig\ref{fig:nf3} again vanish after summing over all fermion loop orientations; in fact, both contain $d_0$: class (j) is proportional to $\cf d_0$, while (k) gives $d_0$.

%
\section{Notation}
\la{se:notation}

To fix our notation: $T^a$ are hermitian generators of a simple Lie algebra, with real and totally antisymmetric structure constants $f^{abc}$ defined by the commutation relation $[T^a,T^b]=i f^{abc}T^c$. 
The Casimir operators of the fundamental and adjoint representations are defined as $T^aT^a=\CF1\!\!1$ and $f^{acd}f^{bcd}=\CA\delta^{ab}$.
We normalize fundamental traces as ${\rm Tr}(T^aT^b)=\TF\delta^{ab}$, denote the number of group generators (gluons) with $\NA$, and the number of quark flavors with $\Nf$. Let us define the following normalized combinations:
\ba
\nf=\frac{\Nf\,\TF}{\CA} \quad,\quad \cf=\frac{\CF}{\CA}\;.
\ea

Higher-order group invariants will enter via traces \cite{vanRitbergen:1998pn}. 
Denoting the generators of the adjoint representation as $[F^a]_{bc}=-if^{abc}$, it is useful to define traces over combinations of symmetric tensors for our discussion:
\ba
&&d_0=\frac{[\str(T^aT^bT^c)]^2}{\NA\TF^2\CA}
\;,\;\;
d_1=\frac{[\str(T^aT^bT^cT^d)]^2}{\NA\TF^2\CA^2}
\;,\;\;
d_2=\frac{\str(T^aT^bT^cT^d)\,\str(F^aF^bF^cF^d)}{\NA\TF\CA^3}
\;,\quad\\&&
d_3=\frac{[\str(F^aF^bF^cF^d)]^2}{\NA\CA^4}
\;,\;\;
d_4=\frac{\str(T^aT^bT^cT^d)\,\str(T^aT^bT^e)\,\str(T^cT^dT^e)}{\NA\TF^3\CA^2}
\;,
\ea
where $\str$ is a fully symmetrized trace (such that $\str(ABC)=\frac12{\rm Tr}(ABC+ACB)$ etc.).

Choosing SU($N$) as gauge group\footnote{For the group U(1) (QED), one simply sets $\CA=\str(F^aF^bF^cF^d)=0$ and $\CF=\TF=\NA=\str(T^aT^bT^cT^d)=1$.} ($\TF=\frac12$, $\CA=N$), some of these normalized invariants read \cite{vanRitbergen:1998pn}
\ba
\label{eq:sun}
\nf=\frac{\Nf}{2N}
\;,\quad
\cf=\frac{N^2-1}{2N^2}
\;,\quad
d_1=\frac{N^4-6N^2+18}{24N^4}
\;,\quad
d_2=\frac{N^2+6}{24N^2}
\;,\quad
d_3=\frac{N^2+36}{24N^2}\;.
\ea
From here, e.g.\ the SU(3) values, corresponding to physical QCD, can be obtained easily.

%
\section{Results}
\la{se:results}

We define the coefficients $b_i$ of the Beta function as
\ba
\partial_{\ln\mu^2}\,a=-a\Big[\ep+b_0\,a+b_1\,a^2+b_2\,a^3+b_3\,a^4+b_4\,a^5+\dots\Big]
\;,\quad
a\equiv\frac{\CA\,g^2(\mu)}{16\pi^2}
\;,
\ea
where $g(\mu)$ is the QCD gauge coupling constant, depending on the regularization scale $\mu$, and we are working in the $\msbar$ scheme, in $d=4-2\ep$ dimensions.
Note that our coupling $a$ is simply a rescaled version of the conventional strong coupling constant $\alpha_s=\frac{g^2(\mu)}{4\pi}$, and that $b_4$ corresponds to 5 loops.
In terms of the normalized color factors introduced in the previous section, we obtain the following result:
\ba
3^1\,b_0 &=& \big[\!-4\big]\nf+11 \;,\\
3^2\,b_1 &=& \big[\!-36\cf-60\big]\nf+102\;,\\
3^3\,b_2&=& \big[132\cf+158\big]\nf^2+\big[54\cf^2-615\cf-1415\big]\nf+2857/2\;,\\
3^5\,b_3&=& \big[1232\cf+424\big]\nf^3
+(150653/2-1188\zeta_3)+432(132\zeta_3-5)d_3
+\\\nonumber
&+& \big[72(169-264\zeta_3)\cf^2+64(268+189\zeta_3)\cf+6(3965+1008\zeta_3)+1728(24\zeta_3-11)d_1\big]\nf^2
+\\\nonumber
&+& \big[11178\cf^3+\!36(264\zeta_3\!-\!1051)\cf^2+\!(7073\!-\!17712\zeta_3)\cf 
+\!3(3672\zeta_3\!-\!39143)+\!3456(4\!-\!39\zeta_3)d_2\big]\nf,\\
\la{eq:result}
3^5\,b_4 &=&\Big[\!-8(107+144\zeta_3)\,\cf+4(229-480\zeta_3)\Big]\,\nf^4
+\Big[c_1\,\cf^2+c_2\,\cf+c_3+c_4\,d_1\Big]\,\nf^3+\dots\;,\\&&
c_1=-6(4961-11424\zeta_3+4752\zeta_4)\;,\quad
c_2=-48(46+1065\zeta_3-378\zeta_4)\;,\\&&
c_3=-3(6231+9736\zeta_3-3024\zeta_4-2880\zeta_5)\;,\quad
c_4=1728(55-123\zeta_3+36\zeta_4+60\zeta_5) \;,
\ea
where $\zeta_s=\zeta(s)=\sum_{n>0}n^{-s}$ are values of the Riemann Zeta function.

For practical reasons, we have performed most of our calculations in Feynman gauge; hence we cannot (yet) claim gauge-parameter cancellation as a check on the result. 
There are, however, a number of other strong checks, as we shall explain now.
First, the coefficients $b_0$ to $b_3$ agree perfectly with the corresponding evaluations up to 4 loops \cite{vanRitbergen:1997va,Czakon:2004bu}, serving as a validation of our whole setup.

Second, already 20 years ago\footnote{The $\cf$ term of \eq\nr{eq:fe} has been known even longer, from QED \cite{PalanquesMestre:1983zy}.}, the leading-order coefficients of the QCD Beta function have been computed in a large $\Nf$ expansion \cite{Gracey:1996he}. In this limit, QCD is equivalent to the non-abelian Thirring model, whose anomalous dimension of $(F_{\mu\nu}^a)^2$ at the $d$-dimensional fixed point gives a result in terms of Gamma functions $\eta(\ep)=\frac{(2\ep-3)\Gamma(4-2\ep)}{16\Gamma^2(2-\ep)\Gamma(3-\ep)\Gamma(\ep)}$. 
The all-order expression for the large-\/$\Nf$ QCD Beta function, written in terms of the coupling $\af=\frac{\Nf\TF g^2(\mu)}{12\pi^2}$, reads \cite{Gracey:1996he}
\ba
\partial_{\ln\mu^2}\af &=& -\af\ep+\af^2+\frac{\af^2}{\nf}\Big(-\frac{11}4+\sum_{j>0}\frac{f_j\,\af^j}{j}\Big)
+{\cal O}\Big(\frac1{\nf^2}\Big) \;,\\
\la{eq:fe}
\sum_{j>0}f_j\,\ep^j &=& -\eta(\ep)
\Big\{4(1+\ep)(1-2\ep)\cf+\frac{4\ep^4-14\ep^3+32\ep^2-43\ep+20}{(1-\ep)(3-2\ep)}\Big\} \;.
\ea
Performing the $\ep$ expansion of \eq\nr{eq:fe}, $f_1$ to $f_4$ agree with the leading-\/$\Nf$ terms of $b_1$ to $b_4$, constituting a first non-trivial check at 5 loops. 

As a third check, the full 5-loop QCD Beta function has recently been obtained for the gauge group SU(3) \cite{Baikov:2016tgj}. 
Reducing our \eq\nr{eq:result} to this special case using \eq\nr{eq:sun} at $N=3$ (ie., $\nf=\Nf/6$, $\cf=4/9$ and $d_1=5/216$), we get
\ba
3^5\,b_4 &\stackrel{\mbox{\tiny SU(3)}}=&
\Big[\frac{1205}{2916}-\frac{152}{81}\,\zeta_3\Big]\,\Nf^4
+\Big[-\frac{630559}{5832} -\frac{48722}{243}\,\zeta_3 +\frac{1618}{27}\,\zeta_4 +\frac{460}{9}\,\zeta_5\Big]\,\Nf^3
+\dots \;,
\ea 
which can be seen to be in full agreement with \cite{Baikov:2016tgj}. 

%
\section{Conclusions}
\la{se:conclu}

We have presented new results for the $\Nf^3$ contribution to the 5-loop QCD Beta function, for the case of a general gauge group. Our methods are highly automated and well suited to evaluate anomalous dimensions and other quantities that can be mapped onto 5-loop massive tadpoles, for which we have high-precision numerical (and partly analytical) results available. Our main result is given in \eq\nr{eq:result}. It agrees favorably with an independent recent calculation \cite{Baikov:2016tgj} that has been performed in parallel to our investigation.

For the special case of the Beta function, an evaluation of the remaining coefficients (proportional to $\Nf^2$, $\Nf^1$ and $\Nf^0$) is under way. While conceptually completely under control, the required computer resources, in particular for the contributions from the gluon propagator, are significantly larger than what was required for the partial result reported here.

Finally, besides the three renormalization coefficients that we have chosen to derive the Beta function, the complete set of anomalous dimensions is within reach. 
A next logical step would be the quark mass anomalous dimension, which is gauge invariant. It is known at 4-loop order for a general gauge group \cite{Chetyrkin:1997dh,Vermaseren:1997fq}, and at 5-loop order for SU(3) \cite{Baikov:2014qja}.
It would also be interesting to keep the gauge parameter (or, at least, linear terms) in the calculations, in order to provide an additional strong check.

%
\acknowledgments

We would like to thank K.~G.~Chetyrkin for discussions, and the authors of~\cite{Baikov:2016tgj} for sharing with us their results prior to publication.
The work of T.L.\ has been supported in part by DFG grants GRK 881 and SCHR 993/2.
A.M.\ is supported by a European Union COFUND/Durham Junior Research Fellowship under EU grant agreement number 267209.
P.M.\ was supported in part by the EU Network HIGGSTOOLS PITN-GA-2012-316704.
Y.S.\ acknowledges support from DFG grant SCHR 993/1, FONDECYT project 1151281 and UBB project GI-152609/VC.
All diagrams were drawn with Axodraw~\cite{Vermaseren:1994je,Collins:2016aya}.

%

\end{document}